\documentclass{vgtc}                          %

\graphicspath{{figures/}{pictures/}{images/}{./}} %

\usepackage{times}                     %

\usepackage{tabu}                      %
\usepackage{booktabs}                  %
\usepackage{lipsum}                    %
\usepackage{mwe}                       %

\usepackage{mathptmx}                  %

\onlineid{0}

\vgtccategory{Research}

\vgtcinsertpkg

\title{Striking the Right Balance: Systematic Assessment of \\ Evaluation Method Distribution Across Contribution Types}

\author{
Feng Lin\thanks{e-mail: flin@unc.edu}\\ %
        \scriptsize UNC Chapel Hill %
\and Arran Zeyu Wang\thanks{e-mail: zeyuwang@cs.unc.edu}\\ %
     \scriptsize UNC Chapel Hill %
\and Md Dilshadur Rahman\thanks{e-mail: dilshadur@sci.utah.edu}\\ %
     \scriptsize University of Utah %
\and Danielle Albers Szafir\thanks{e-mail: danielle.szafir@cs.unc.edu}\\ %
     \scriptsize UNC Chapel Hill %
\and Ghulam Jilani Quadri\thanks{e-mail: quadri@ou.edu}\\ %
     \scriptsize University of Oklahoma
     }

\abstract{%
In the rapidly evolving field of information visualization, rigorous evaluation is essential for validating new techniques, understanding user interactions, and demonstrating the effectiveness and usability of visualizations. Faithful evaluations provide valuable insights into how users interact with and perceive the system, enabling designers to identify potential weaknesses and make informed decisions about design choices and improvements.
However, an emerging trend of multiple evaluations within a single research raises critical questions about the sustainability, feasibility, and methodological rigor of such an approach. New researchers and students, influenced by this trend, may believe -- multiple evaluations are necessary for a study, regardless of the contribution types. However, the number of evaluations in a study should depend on its contributions and merits, not on the trend of including multiple evaluations to strengthen a paper. So, how many evaluations are enough? This is a situational question and cannot be formulaically determined. Our objective is to summarize current trends and patterns to assess the distribution of evaluation methods over different paper contribution types. In this paper, we identify this trend through a non-exhaustive literature survey of evaluation patterns in 214 papers in the two most recent years' VIS issues in IEEE TVCG from 2023 and 2024.
We then discuss various evaluation strategy patterns in the information visualization field to guide practical choices and how this paper will open avenues for further discussion.
}

\CCScatlist{
  \CCScatTwelve{Human-centered computing}{Visu\-al\-iza\-tion}{Visualization design and evaluation methods}{}
}

\usepackage{colortbl}
\usepackage{color}
\usepackage{tabularray}

\begin{document}
\maketitle

\section{Introduction}
\label{sec-intro}

Evaluating visualization system, design, and technique is fundamental to ensuring the effectiveness, usability, and impact of visualizations in practice~\cite{borgo2018information, carpendale2008evaluating}. Evaluation enables researchers to identify potential issues or weaknesses in visualizations and allows designers to address these limitations before the system is released or the design is implemented. Faithful evaluation provides valuable predictive insights into how users interact with and perceive the system, enabling designers to make informed decisions about design choices and improvements. Efficient evaluations can also serve as a mechanism to demonstrate the system's merits, which is crucial for gaining support, funding, and prototype implementation. Three primary types of evaluation are typically employed in visualization research: \textit{quantitative}, \textit{qualitative}, and \textit{case studies}. Qualitative methods, such as interviews, observations, and focus group workshops, offer in-depth understandings of users' experiences and perceptions~\cite{quadri2024do}. Quantitative methods, such as metrics-based assessments and graphical perception experiments, provide quantified data for comparisons and generalizations~\cite{tseng2023evaluating}.
Case studies typically provide illustrative usage cases focusing on application insights for prototype demonstrations~\cite{smart2019color}.
Lam et al.\cite{lam2011empirical} conducted a comprehensive literature survey and designed a taxonomy for choosing evaluation methods according to specific research goals. They also found about half of the sample papers from 1995 to 2011 did not utilize any systematic evaluation.
However, recent years have seen a steady growth in the number of evaluations employed within a given paper, due in part to the adoption of crowdsourced platforms, as highlighted by Borgo et al.~\cite{borgo2018information}.

While different forms of evaluation may complement one another to provide a more holistic perspective on effectiveness, there is an emerging trend that increasing the number of evaluations within a study would always be beneficial, raising important questions about the sustainability, feasibility, and methodological rigor of this approach and increasing pressure on new evaluation studies~\cite{greenberg2008usability}. 
Adding more evaluations may offer diminishing returns.
For instance, an application paper may employ various evaluation forms (e.g., quantitative, qualitative, and case studies) when a single evaluation might suffice to validate usability, performance, and task accuracy. This trend can lead new researchers and students to believe that multiple evaluations are necessary for a study. However, the number of evaluations in a study should depend on its contributions and their alignment with the evaluation and not simply adding more evaluations for the sake of more evaluation.

Determining the optimal number of evaluations for a given system, technique, application, or experiment is a challenging task that requires careful consideration of various factors, including the visualization's scope, related design complexity, intended users, target tasks, study design, and available resources. 
To better tackle this challenge, we explored the general patterns of evaluation methods utilized in different types of papers and proposed general recommendations for researchers.

\begin{figure*}[ht]
\centering
\includegraphics[width=1.95\columnwidth]{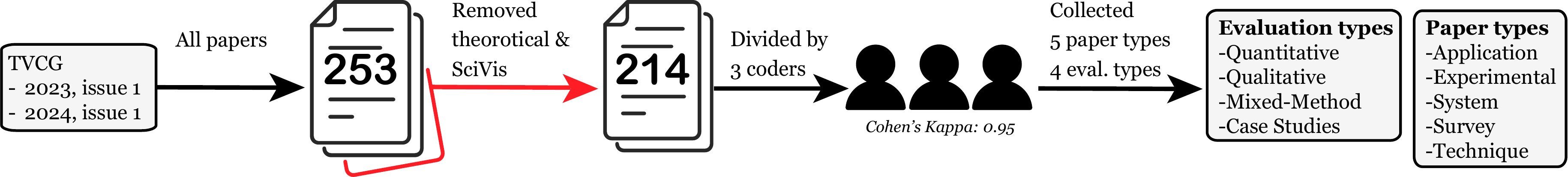}
\caption{
An overview of our literature selection process, which identified 214 papers from IEEE TVCG 2023 Issue 1 and 2024 Issue 1. 
Three coders qualitatively coded them into four evaluation types and five paper types with a high agreement.
}
\label{fig:literaturesurvey}
\end{figure*}

Our objective is to catalyze a conversation on how many evaluations should be designed and conducted in visualization studies. However, the question---\textit{"How many evaluations are enough?"} is situational and cannot be formulaically determined. 
While past methodological analyses have focused on theoretical alignment between method and contribution type, we aimed to summarize current trends and patterns of evaluations in visualization research to understand high-level norms and practices researchers are employing across different types of papers and their contributions. 
We explored 
evaluation patterns in 214 papers in the two most recent years' VIS issues in IEEE TVCG categorized by 5 types of papers (see~\autoref{fig:literaturesurvey})---\emph{application}, \emph{experimental}, \emph{system}, \emph{survey}, and \emph{technique}---and 3 types of evaluations---\emph{quantitative}, \emph{qualitative}, and \emph{case study}, with \emph{mixed-method} identified as a more specific form of analysis. 
We learned that the visualization study with the optimal number and type of evaluation should prove whether a solution from the given visualization system, technique, application, or experiment meets the research goals.

Built upon the preliminary analysis of 214 papers, we propose that different papers \textit{types} tend to require tailored evaluation approaches---experimental papers could employ both quantitative and qualitative evaluations; survey papers may only need a single qualitative study; system papers could incorporate multiple case studies and qualitative evaluations; technique papers may rely on multiple quantitative evaluations and case studies; and application papers prioritize multiple case studies to demonstrate real-world utility. 
However, researchers must consider limited resources and time when choosing evaluation methods to ensure validation without overextending resources.
For example, researchers often conduct multiple evaluations for a single system, technique, or experiment paper, with 35.05\% using a combination of three evaluation types and 36.92\% using a combination of two types. 
We further reported differences across paper types. For example, a large portion of survey papers do not employ evaluation methods since their emphasis is on literature review and synthesis rather than on applying evaluation techniques to validate models and algorithms.

However, it remains unclear whether the insights gained 
from combining multiple evaluations
justify the resources invested and if fewer evaluations could be sufficient. 
The long-term viability and sustainability of these strategies also remain speculative. 
Therefore, 
we propose that a more concrete principle for evaluation choices in visualization research is dependent on the specific contribution of the study, which further remains a challenging problem.
Addressing this challenge will require a more holistic perspective on evaluation, drawing from both practice and theory, including expert perspectives and metamethodological studies.

\section{Related Work}
\label{sec-background}

Past work offers multiple perspectives on evaluation approaches. Early studies categorized evaluation techniques into four main areas: usability evaluation, controlled experiments comparing design attributes, controlled experiments comparing multiple visualizations, and case studies\cite{komlodi2004information}. Lam et al.\cite{lam2011empirical}'s taxonomy of task-based evaluation methods emphasized the importance of aligning evaluation types with specific research goals. They reviewed 850 papers and categorized seven evaluation scenarios to be used for researchers to "reflect on evaluation goals before choosing evaluation methods." Borgo et al.\cite{borgo2018information} categorized 190 crowdsourcing experiments reported in 82 papers into six main aspects according to task type, reproducibility, and validity.
However, as the field has matured, the methods employed in evaluation have subsequently expanded, offering a wealth of new perspectives on visualization effectiveness \cite{wall2022vishikers}. 

Evaluation practices have significantly evolved over the years. Initially, many studies lacked standard evaluation frameworks and often provided little to no formal evaluation. Lam et al. found that over half of the sample papers (i.e., 489 out of 850) from major visualization venues during 1995-2011 did not use any evaluation~\cite{lam2011empirical}. Recently, there has been a shift towards more systematic and varied evaluation methodologies in the field.
Isenberg et al. \cite{isenberg2013systematic} conducted a similar review of 581 IEEE Visualization conference papers from 1997-2012 and found that about 97\% of papers applied at least one type of evaluation. They also identified a growing trend in the IEEE Information Visualization conference (and subsequently within IEEE VIS after its unification) of using human subjects evaluation methods, such as user performance and work practice assessments. Crowdsourcing's ability to reduce barriers to participant recruitment further accelerated this trend \cite{borgo2018information}.

Adapting to an expanded set of evaluation methods presents several challenges. For example, Carpendale \cite{carpendale2008evaluating} identified three critical factors in evaluation studies: generalizability, precision, and realism, which often conflict with each other. Quantitative evaluations typically offer high precision and generalizability because they use numerical data and statistical metrics to assess visualization performance. However, they often lack realism due to the controlled and artificial nature of study conditions. Conversely, qualitative evaluations tend to provide more realism by focusing on real-life interactions through detailed observations, interviews, and comprehensive feedback analysis. Yet, they lack precision due to their subjective nature and the difficulty in replicating results.

Evaluation methods may suffer from additional challenges, such as aligning visualization objectives with experimental studies.
The ill-defined study tasks systematically evaluate knowledge acquisition while relying only on low-level tasks in the experiment design highlighting the validation complexity ~\cite{munzner2009nested}.
Recent efforts aim to make it easier to assess trade-offs in evaluation. 
For example, six levels of the knowledge acquisition framework adapted from Bloom's taxonomy helped researchers systematically evaluate a visualization's capabilities \cite{burns2020evaluate}. 
Designing studies to complement past evaluations, such as assessing qualitative high-level comprehension to analyze the utility of lower-level quantitative graphical perception studies, can offer new perspectives on past guidance~\cite{quadri2024do}. 
Sperrle et al. introduced design dimensions for structured evaluation and identified key factors that influence interpretability and explainability~\cite{sperrle2021survey}. 
They emphasized the importance of open-sourcing evaluation materials but also called for the development of a structured evaluation framework that enables comparable evaluations and balances between over-evaluation and insufficient evaluation.

The choice of evaluation methods in visualization systems is critical, with each approach having specific trade-offs that may work in complement with one another. 
However, excessive evaluations can lead to resource inefficiency, while insufficient evaluation may compromise reliability and decision-making. 
Buxton and Greenberg\cite{buxton2010sketching, greenberg2012sketching,greenberg2008usability} call for careful reflection on the evaluation practices and the importance of finding a balance between the number of evaluations and the time or costs for it, arguing that informed design is essential for innovation but sometimes the best evaluation is no evaluation, harkening back to earlier trends in visualization observed by Lam et al. \cite{lam2011empirical}. 
Achieving a balanced approach to evaluation is essential for the field, ensuring research efforts are 
aligned with research goals and available resources.

\section{Methodology}
\label{sec-method}
\definecolor{myLightBlue}{rgb}{0.68, 0.85, 0.90}
\definecolor{myOrange}{rgb}{1, 0.65, 0.3}
\definecolor{myLightGreen}{rgb}{0.56, 0.93, 0.56}
\definecolor{myLightred}{rgb}{1, 0.5, 0.56}

\begin{table}[]
\caption{Table serves as a template that outlines the types of research papers (e.g., Application, Experimental, System, Survey, Technique) and their associated evaluation methods which we utilized in the review process. The columns "Quantitative", "Qualitative", "Mixed-Method" (i.e., Mixed), and "Case Study" indicate the number of specific evaluation methods used: None - this evaluation is not used, Single - one evaluation of this type, Multiple - multiple evaluations of this type, and Yes/No for mixed-methods.}
\centering
\small
\begin{tabular}{|c|c|>{\columncolor{myLightBlue}}c|>{\columncolor{myOrange}}c|>{\columncolor{myLightGreen}}c|>
{\columncolor{myLightred}}c|}
\hline
\textbf{Paper} & \textbf{Type} & \textbf{Quantitative} & \textbf{Qualitative} & \textbf{Case Study} & \textbf{Mixed} \\ \hline

\cite{epperson2023dead}         &   System   &        Single      &      Multiple       &     Multiple       &     Yes         \\ \hline

   \cite{heer2023mosaic}      &   Technique   &        Single      &      None       &     Multiple       &     No         \\ \hline
   
    \cite{moritz2023average}     &  Experiment    &    Multiple          &        Multiple     &   None         &     No     \\ \hline
    
\cite{oral2023information}         &  Survey    &         None     &       None      &    None        &     No         \\ \hline

  \cite{shen2022idlat}       &   Application   &      Single        &      Single       &   Multiple         &       Yes       \\ \hline

 ...      &   ...   &      ...        &      ...       &   ...        &       ...     \\ \hline
\end{tabular}
\label{figs:table-0}
\end{table}

We conducted a review of literature from IEEE Transactions on Visualization and Computer Graphics (TVCG), specifically focusing on Issue 1 from 2023 and 2024 (i.e., issues of full papers in IEEE VIS 2023 and 2024). Our non-exhaustive survey categorized and analyzed evaluation methods employed in these papers to provide insights into approaches for evaluation count in a given study. 
We initially collected 253 papers from Issue 1 of IEEE TVCG for 2023 and 2024. To maintain a clear focus on our specific area of interest, we systematically excluded papers not directly related to information visualization, such as those on scientific visualization and theoretical studies (e.g., \cite{morrical2022quick,ageeli2022multivariate}), resulting in a final selection of 214 papers. 
We focused on papers classically considered information visualization as these papers would allow direct comparison against past surveys of evaluation techniques to understand patterns over time.
We then categorized the evaluation methods into three groups (see~\autoref{figs:table-0}): \textit{Quantitative} methods, \textit{Qualitative} methods, and \textit{Case Studies}. We also considered papers utilizing both quantitative and qualitative methods in any single study as \textit{Mixed Methods}.

According to Munzner \cite{munzner2008process}, 
"selecting a target paper type in the initial stage can avert an inappropriate choice of validation methods" so we further grouped surveyed papers into paper types (see \cref{figs:table-0}):

\begin{itemize}
\item \textit{Application} papers present novel systems, demonstrating their utility in practical applications employing studied techniques from visualization and related fields~\cite{fernandez2022ergoexplorer,hoque2022visual}.

\item \textit{Experimental} papers provide empirical evidence on visualization methods' performance and user interaction through user studies 
or classical experiments~\cite{lee2022affective, padilla2022multiple}.

\item \textit{Survey} papers offer an inclusive view of a research domain, summarizing existing literature and research trends and suggesting future directions~\cite{mcnutt2022no,panagiotidou2022communicating}.

\item \textit{System} papers detail the design, implementation, and evaluation of new systems composed of different models, algorithms, and technologies that support visualization tasks~\cite{vajiac2022trafficvis, yu2022pseudo}.

\item \textit{Technique} papers introduce novel methods, explaining their advantages and use cases, often including state-of-the-art comparative evaluations~\cite{swift2022visualizing,han2022sizepairs}.

\end{itemize}

More detailed definitions and analysis of these 5 paper types are provided in~\cref{sec-results}. Surveyed papers can be accessed at  \href{https://docs.google.com/spreadsheets/d/1tgKlUm23xsazU_rFGVi_c8Kkq-fVwVnm5_G4CgHOv88/edit?gid=0#gid=0}{here}.
 
Quantitative evaluations \textit{measure} the level of effectiveness and efficiency of visualization systems. These methods are useful for producing objective and reproducible results that can be statistically analyzed. Specific examples of quantitative methods include measuring user performance metrics, such as accuracy and error rates\cite{rodrigues2022relaxed, ha2022unified, patil2022studying}. Heatmaps and eye-tracking technology provide visual representations of user interactions and gaze patterns, highlighting areas of frequent activity and identifying navigation patterns \cite{ha2022unified, morariu2022predicting,shin2022scanner}. Task completion time and survey scores, often collected through standardized usability questionnaires, determine how quickly and correctly users can perform tasks using the visualization \cite{tandon2022measuring, holder2022dispersion, morariu2022predicting}. These quantitative methods collectively offer a variety of perspectives for evaluating the performance and usability of information visualization systems.

Qualitative evaluations focus on understanding user experiences, feedback, and perceptions. These methods help researchers understand how users interact with visualizations in real-life contexts. They can uncover
potential usability issues that may not be evident through quantitative metrics alone. Examples of qualitative methods include user interviews \cite{zong2022animated}, expert feedback \cite{vajiac2022trafficvis}, open-ended surveys \cite{linhares2022largenetvis}, and think-aloud protocols \cite{warchol2022visinity}. %
 
Case study methods involve in-depth examinations of how visualization systems can be applied in real-world applications to solve specific problems or achieve particular goals. For example, Yu et al. \cite{yu2022pseudo} did an in-person case study to show their model PSEUDo’s usability in a real-world use case in the energy domain with expert feedback. Yuan et al. \cite{yuan2022visual} conducted two case studies with different large architectural spaces to demonstrate the effectiveness of ArchExplorer. 

Mixed methods combine qualitative and quantitative approaches in one single study to provide a more comprehensive evaluation of visualization systems. Unlike approaches that simply utilize quantitative and qualitative methods in separate studies of a paper, mixed methods should deliberately combine these methods within the same study
\cite{almeida2018strategies, small2011conduct}. By applying the strengths of both types of evaluations, mixed methods can offer a balanced perspective on usability from the users' perspective and effectiveness from the objective performance perspective. For instance, a study might use a controlled experiment to gather quantitative performance metrics and follow it up with user interviews reflecting on the experimental tasks to explore the users' perspectives. For example, Ghai et al. \cite{ghai2022d} conducted a user study and evaluated their tools from both quantitative methods (i.e., analyzing utility metrics and bias metrics) and qualitative methods (i.e., analyzing direct quotes from participants).
Li et al. \cite{li2022dual} conducted a mixed methods study; first utilized quantitative evaluation to compare the performance of their method with state-of-the-art methods on time cost and the five other metrics and then they applied qualitative methods as a more intuitive perspective by showing the outputs of several representative datasets and the improvements they made in three applications.

Additionally, for papers utilizing quantitative, qualitative, and case study methods, we further analyzed whether these studies employed \textit{single} or \textit{multiple} evaluation approaches. This deeper examination helped us quantify the prevalence of diverse evaluation strategies within the field.
This approach allowed us to also understand how the number of different evaluations applied to any single contribution has changed over time. Based on our grouping of the 214 papers into four types of evaluation and five contribution types, we can begin to illustrate the current trends in the types and frequency of evaluations used in modern visualization research.

\section{Results}
\label{sec-results}
\begin{figure}[]
\centering
\includegraphics[width=1\columnwidth]{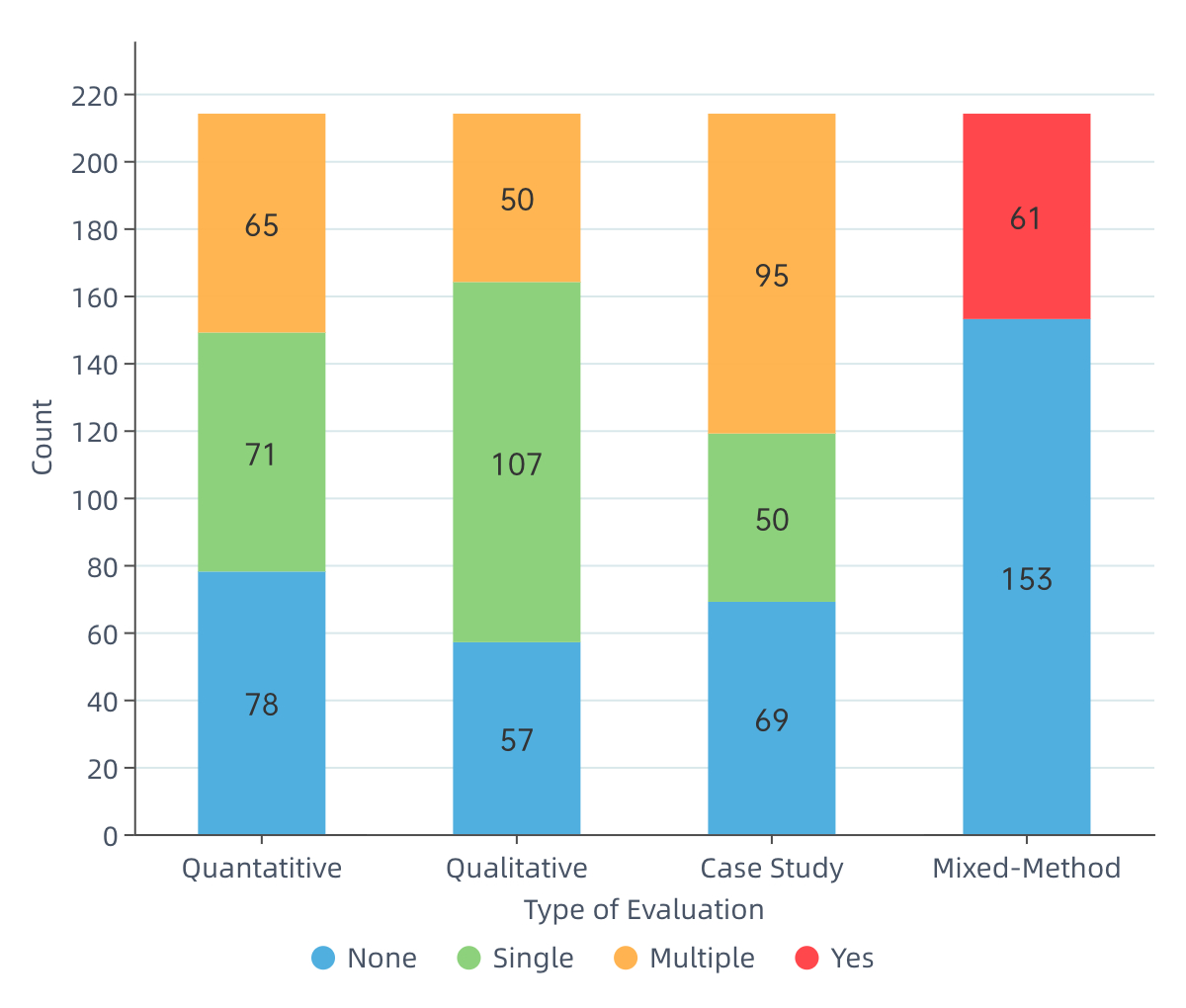}
\caption{Distribution across different sub-types of four evaluation methods- quantitative, qualitative, case study, and mixed methods for 214 papers. \textit{None}---not utilized for a given study, \textit{single}---one evaluation under this type, \textit{multiple}---multiple evaluations under this type, and \textit{Yes} and \textit{No} only for mixed methods.}
\vspace{-1em}
\label{figs:1}
\end{figure}

\subsection{Evaluation Types}
We classified these evaluation types based on the evaluation frequency per paper into \textit{none}, \textit{single}, and \textit{multiple}. For mixed methods, we identified whether the method was absent (\textit{none}) or present (\textit{yes}). Distinct patterns of evaluation counts are illustrated in~\autoref{figs:1}.

Our analysis demonstrated that 63.55\% of the reviewed papers include at least one quantitative evaluation, with a substantial number (30.37\%) employing multiple quantitative methods. This underscores the importance of quantitative evaluation in the field of information visualization. However, 36.45\% of the papers did not employ any quantitative evaluation methods. This trend can be attributed to specific studies where conducting quantitative evaluations may be challenging or less relevant. 
Survey papers (e.g.,~\cite{oral2023information}) often offer theoretical contributions without objective measurements. New visualization systems~\cite{batch2023wizualization,elavsky2023data} solve problems, meaning they have no prior system to compare against. Instead, these studies focus on synthesizing existing knowledge or rely on qualitative feedback and real-world case studies for evaluation.

We found that 73.36\% of the reviewed papers utilize one or more qualitative evaluation methods, indicating a strong focus on detailed user experiences and expert feedback to validate the performance and subjective usability of visualization systems. However, papers tend to include only one qualitative evaluation (50\%) compared to multiple qualitative methods (23.36\%). This discrepancy may be attributed to the resource-intensive nature of qualitative studies. Gathering and analyzing qualitative data, such as through interviews, focus groups, or observational studies, requires significant time and effort. As a result, researchers often prioritize a single, well-chosen qualitative method that provides in-depth insights into user experiences and contextual nuances. While using multiple qualitative methods would offer richer data, it also demands more resources, which can be a limiting factor, especially in studies with constrained budgets and timelines.

Only 28.50\% of the studies employ mixed method evaluations. The relatively lower usage of mixed methods 
may be due to the complexity of designing evaluations that integrate both quantitative and qualitative approaches within a single study. Unlike approaches that simply apply quantitative and qualitative methods in separate studies of one paper, true mixed method evaluations involve the deliberate combination of these methods within the same study context. This integration allows for a more comprehensive analysis by capturing both numerical data and in-depth user insights, providing a richer understanding of the subject under investigation. However, because this approach requires careful coordination and alignment of different evaluation strategies, it is often reserved for specific and nuanced cases where both types of data are essential for drawing meaningful conclusions (e.g., Padilla et al. \cite{padilla2022multiple}). This may explain why mixed-method evaluations, while powerful, are less commonly employed despite their potential benefits.

Our results also demonstrated a strong preference for using multiple case studies (44.39\%), highlighting the value of validating visualization systems and tools across various real-world scenarios. This approach may more directly illustrate a system's practical relevance and generalizability.

Based on our analysis, the use of evaluation methods in information visualization research varies significantly depending on the context and objectives of each study. A substantial number of papers (63.55\%) include at least one quantitative method, while an even larger number (73.36\%) utilize at least one qualitative method. This underscores the critical role that both quantitative and qualitative evaluations play in the field. Additionally, mixed method evaluations and case studies, though less frequently used, provide valuable insights and enhance the generalizability and practical relevance of visualization systems. 

\begin{figure}[]
\centering
\includegraphics[width=0.75\columnwidth]{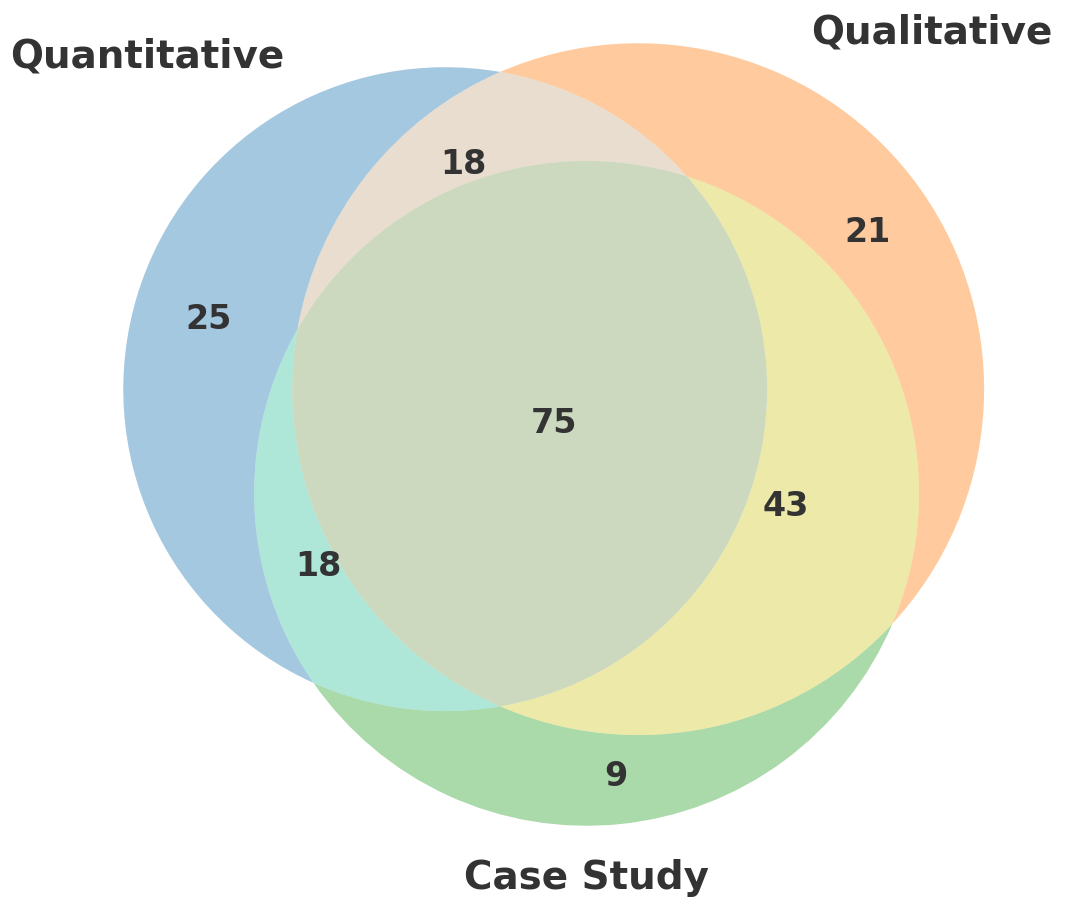}
\caption{Venn diagram of the distribution for different evaluation types used in 214 papers. The overlapped sections represent studies that utilize multiple evaluation methods, including quantitative, qualitative, and case study methods.}
\vspace{-1em}
\label{figs:table2}
\end{figure}

\paragraph{Evaluation using Comprehensive Approach:}
We utilized a Venn Diagram (see~\autoref{figs:table2}) to illustrate the distribution and overlap of the three (quantitative, qualitative, and case study) primary evaluation methods for 214 papers.  
Approximately 35.05\% of 214 papers employed all primary three evaluation methods. This significant overlap indicates research embracing a comprehensive approach to evaluation, capturing both quantitative metric data and user experiences while demonstrating practical relevance through real-world applications. The use of the three primary methods provides a robust and well-rounded evaluation, ensuring findings are validated from multiple perspectives. However, this approach might consume excessive resources.
Only 9 papers utilized case studies alone, while the remaining case studies were accompanied by other evaluation methods (18 with quantitative and 43 with qualitative). We argue that case studies alone may lack depth in empirical analysis or user experience data, limiting their generalizability beyond the target application. However, by combining case studies with other methods, researchers offer a holistic evaluation addressing various facets of visualization effectiveness, from usability and user satisfaction to technical performance and scalability.

Additionally, 43 studies utilized only qualitative methods and case studies, emphasizing the importance of user experience, contextual relevance, and practical application in information visualization. While quantitative evaluations provide valuable performance data, the subjective and context-dependent nature of visualizations often necessitates a more nuanced approach. Qualitative feedback and real-world case studies offer the depth and flexibility needed to validate the effectiveness of visualizations in diverse and dynamic environments. 
Although the choice of evaluation methods should align with the research objectives, combining multiple evaluation methods can 
typically enhance the research by delivering a more comprehensive and inclusive analysis. To investigate the patterns of evaluation frequency further, we extended our analysis to different paper types.

\begin{figure}[]
\centering
\includegraphics[width=1\columnwidth]{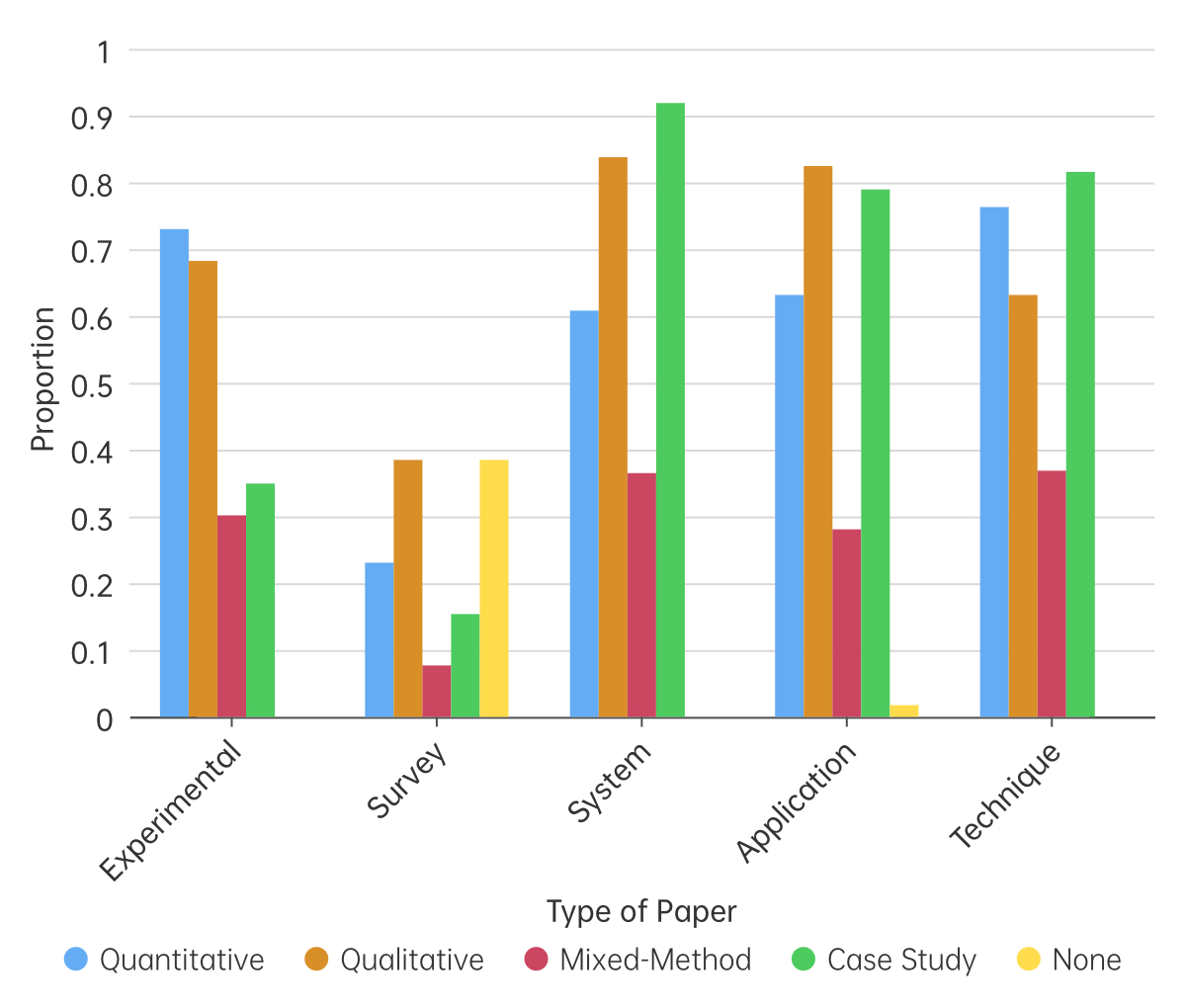}
\caption{Bar chart of the proportion (0-1) for the five categories of papers, including experimental, survey, system, application, and technique (survey paper has only 13 examples). The grouped bar chart demonstrates the distribution of different evaluation types on each paper category. For example, quantitative and case studies are heavily employed for technique-based papers whereas experimental papers have more quantitative and qualitative evaluations}
\vspace{-1em}
\label{figs:table4}
\end{figure}

\subsection{Evaluation per Paper Types}
The evaluation frequency varies with the type of contribution each paper makes. Consequently, we further investigated the trend in mapping five paper types to evaluation methods.

As illustrated in~\autoref{figs:table4}, ~73.02\% of experimental papers utilized quantitative methods and 68.25\% of them employed qualitative methods. The main objective of this combination of both quantitative metric data and 
qualitative perspectives could be conducting richer data analysis, allowing quantitative data to address hypotheses and qualitative data to explain or contextualize observed differences. Such strategies allow for a comprehensive evaluation of hypotheses, assess the effectiveness of new visualization methods, and enhance the reliability of the research conclusions through a more nuanced evaluation.

Of the 13 survey papers, 5 employed no evaluation method, while another 5 used qualitative methods. This indicates a tendency in survey papers to either forgo formal evaluation or favor qualitative approaches as in interview studies.
Many survey papers do not include any evaluation methods at all since they focus on summarizing existing research and providing holistic overviews of the research domains. These findings suggest that survey papers prioritize outlining the research problem over experimental validation.

91.89\% of system papers incorporate case studies to demonstrate the practical 
utility and effectiveness of their proposed systems. This high percentage underscores the critical role of case studies in providing concrete evidence of a system's utility and effectiveness in real-world applications. By showcasing how a system performs in specific contexts, researchers can better communicate the value and potential impact of their work. 83.78\% of system papers also apply qualitative methods to guide iterative design improvements and ensure whether the system meets the needs of its intended users. 

Similar patterns are noted in application papers where qualitative methods (82.46\%) and case studies (78.95\%) are prioritized. Application papers focus on real-world usage, so concrete case studies and qualitative evaluations from users’ perspectives are 
needed to show the core objectives of an application are addressed. 
Technique papers 
rely heavily on quantitative evaluations (76.32\%) and case studies (81.58\%) reflecting technical validity and performance improvements are necessary for faithfully evaluating a newly proposed technique.

Furthermore, it is worth noting that the mixed method evaluation is employed across three major paper types (i.e., experimental, system, and technique), with about 30-36\% of each type utilizing this approach. We argue that a conclusive paper often needs to evaluate multiple aspects regardless of its paper type. Mixed method evaluations that can include multiple factors such as usability, performance, accuracy, user satisfaction, and real-world applicability in one study, helping more comprehensively assess a solution using a single study. 
However, as we discussed in Section 4.1, the design of mixed methods evaluations is complex and requires careful planning and execution to balance and integrate different evaluation dimensions effectively.

\section{How Many Evaluations are Enough?}
\label{sec-discussion}

We conducted a non-exhaustive survey to understand current patterns in visualization evaluation methods and frequency.
While this question is inherently situational and cannot be answered directly, we provide high-level recommendations based on the summarized trends and patterns of evaluations in information visualization research.

The optimal number and types of evaluations for visualization studies 
should be sufficient to prove whether a solution from a given visualization system, technique, application, or experiment meets the research goals. Our survey and review reveal that the evaluation frequency pattern varies with the paper's specific contribution. 
The right level of evaluation may range from a single detailed study to multiple case studies. Furthermore, it may involve examining different independent facets of the research's goals, such as assessing the performance of a technique through a quantitative study and evaluating usability using a qualitative study. 
Experimental papers typically use multiple evaluations, both quantitative and qualitative, for thorough validation. Survey papers often need no evaluation but qualitative evaluation can be added to externally confirm the core synthesis. System and technique papers benefit from a mix of case studies and quantitative evaluations. Application papers tend to need multiple qualitative evaluations and case studies for practical insights.

Evaluation strategies should be context-dependent~\cite{lam2011empirical, borgo2018information}. While rigorous formal evaluations are indispensable for validating specific hypotheses or measuring performance, other situations may benefit from more flexible, iterative, and exploratory approaches. By tailoring evaluation methods to the unique objectives and stages of each study, researchers can ensure that their assessments are both effective and conducive to advancing the field of information visualization. 
Additionally, not every study necessitates an evaluation\cite{greenberg2008usability}. In some cases, the most appropriate approach might be to forgo formal evaluations altogether. This perspective is supported by arguments suggesting that user studies can sometimes be counterproductive or even harmful~\cite{buxton2010sketching,greenberg2012sketching}. They argue that rigidly structured evaluations might constrain creative exploration and innovation, particularly in the early stages of design. 
However, the choice not to formally evaluate a contribution should be done with care---conclusions that seem obvious in the moment may not hold in practice.

Based on our analysis from 214 papers in \cref{sec-results}, we recommend experimental papers use both quantitative and qualitative evaluations, often with mixed methods, and typically need multiple evaluations to validate hypotheses comprehensively. 
Survey papers usually do not include any formal evaluation but may include a single qualitative evaluation study to understand expert opinions or assess the comprehensiveness of the literature review. 
System papers typically incorporate multiple case studies and at least one qualitative evaluation to elicit expert perspectives on a system. These papers need to demonstrate practical implementation and effectiveness, requiring contextual user feedback and real-world applications. 
Technique papers, on the other hand, can rely on multiple quantitative evaluations and case studies to establish technical validity and performance improvements, ensuring robustness and applicability. 
Application papers can heavily utilize case studies: given their focus on real-world usage, demonstrations of utility in specific contexts are necessary.

Determining the optimal number and type of evaluations for a study is crucial and complex. It depends on resources, time, practical application, and the study's objectives. While thorough evaluation is essential for validating visualization techniques, the methodological rigor is often influenced by the study's specific goals and \textit{study type}.
Research should consider how well each evaluation outcome aligns with the goals of the research, balancing coverage of these goals with available resources. 
Research must account for limited resources and time constraints when choosing evaluation methods to ensure meaningful validation without overextending their resources.

\section{Conclusion}
\label{sec-conclusion}

Rigorous evaluation is crucial for validating new visualization techniques, understanding user interactions with visualizations, and demonstrating the effectiveness of visualization systems. Our non-exhaustive review of recent literature reveals a strong preference for combining multiple evaluation methods (e.g., quantitative, qualitative, case studies, mixed-method) within a single study. However, we contend that using multiple evaluations may not always be necessary or practical and can sometimes be excessive in terms of time and resources. Our takeaways emphasize that the selection of evaluation methods should be driven by research goals and the nature of the contribution, rather than merely aiming to strengthen the study, ensuring robust and reliable evaluations.

A notable limitation of our study is the absence of a temporal analysis. While we have focused on papers from 2023-2024, our study does not account for how evaluation practices and methodologies may have evolved. A more comprehensive temporal analysis could have provided insights into trends, shifts in evaluation approaches, and how these changes correlate with broader developments in the field of information visualization. Another critical limitation is the assumption that increased evaluation work correlates with improved objective quality of contributions. While we observed a rise in the quantity and complexity of evaluations conducted in recent papers when compared to past surveys, it is not necessarily indicative of an improvement in the fundamental quality or innovation of the contributions themselves. The trend suggests the need for future research to examine whether the increase in evaluation methods truly results in better outcomes or if it simply reflects a shift in academic expectations without a corresponding improvement in quality.

\acknowledgments{%
We thank the reviewers for their insightful comments.
This work was supported by the National Science Foundation under grant No.2127309 to the Computing Research Association for the CIFellows project and NSF IIS-2046725.
}

\bibliographystyle{abbrv-doi}

\bibliography{template}

\begin{thebibliography}{10}

\bibitem{ageeli2022multivariate}
A.~Ageeli, A.~Jaspe-Villanueva, R.~Sicat, F.~Mannuss, P.~Rautek, and M.~Hadwiger.
\newblock Multivariate probabilistic range queries for scalable interactive 3d visualization.
\newblock {\em IEEE Transactions on Visualization and Computer Graphics}, 29(1):646--656, 2022.

\bibitem{almeida2018strategies}
F.~Almeida.
\newblock Strategies to perform a mixed methods study.
\newblock {\em European Journal of Education Studies}, 2018.

\bibitem{batch2023wizualization}
A.~Batch, P.~W. Butcher, P.~D. Ritsos, and N.~Elmqvist.
\newblock Wizualization: A “hard magic” visualization system for immersive and ubiquitous analytics.
\newblock {\em IEEE Transactions on Visualization and Computer Graphics}, 2023.

\bibitem{borgo2018information}
R.~Borgo, L.~Micallef, B.~Bach, F.~McGee, and B.~Lee.
\newblock Information visualization evaluation using crowdsourcing.
\newblock In {\em Computer Graphics Forum}, vol.~37, pp. 573--595. Wiley Online Library, 2018.

\bibitem{burns2020evaluate}
A.~Burns, C.~Xiong, S.~Franconeri, A.~Cairo, and N.~Mahyar.
\newblock How to evaluate data visualizations across different levels of understanding.
\newblock In {\em 2020 IEEE Workshop on Evaluation and Beyond-Methodological Approaches to Visualization (BELIV)}, pp. 19--28. IEEE, 2020.

\bibitem{buxton2010sketching}
B.~Buxton.
\newblock {\em Sketching user experiences: getting the design right and the right design}.
\newblock Morgan kaufmann, 2010.

\bibitem{carpendale2008evaluating}
S.~Carpendale.
\newblock Evaluating information visualizations.
\newblock pp. 19--45, 2008.

\bibitem{elavsky2023data}
F.~Elavsky, L.~Nadolskis, and D.~Moritz.
\newblock Data navigator: an accessibility-centered data navigation toolkit.
\newblock {\em IEEE Transactions on Visualization and Computer Graphics}, 2023.

\bibitem{epperson2023dead}
W.~Epperson, V.~Gorantla, D.~Moritz, and A.~Perer.
\newblock Dead or alive: Continuous data profiling for interactive data science.
\newblock {\em IEEE Transactions on Visualization and Computer Graphics}, 2023.

\bibitem{fernandez2022ergoexplorer}
M.~M. Fern{\'a}ndez, S.~Rado{\v{s}}, K.~Matkovi{\'c}, M.~E. Gr{\"o}ller, and C.~Delrieux.
\newblock Ergoexplorer: Interactive ergonomic risk assessment from video collections.
\newblock {\em IEEE Transactions on Visualization and Computer Graphics}, 29(1):43--52, 2022.

\bibitem{ghai2022d}
B.~Ghai and K.~Mueller.
\newblock D-bias: A causality-based human-in-the-loop system for tackling algorithmic bias.
\newblock {\em IEEE Transactions on Visualization and Computer Graphics}, 29(1):473--482, 2022.

\bibitem{greenberg2008usability}
S.~Greenberg and B.~Buxton.
\newblock Usability evaluation considered harmful (some of the time).
\newblock In {\em Proceedings of the SIGCHI conference on Human factors in computing systems}, pp. 111--120, 2008.

\bibitem{greenberg2012sketching}
S.~Greenberg, S.~Carpendale, N.~Marquardt, and B.~Buxton.
\newblock {\em Sketching user experiences: The workbook}.
\newblock Elsevier, 2012.

\bibitem{ha2022unified}
S.~Ha, S.~Monadjemi, R.~Garnett, and A.~Ottley.
\newblock A unified comparison of user modeling techniques for predicting data interaction and detecting exploration bias.
\newblock {\em IEEE Transactions on Visualization and Computer Graphics}, 29(1):483--492, 2022.

\bibitem{han2022sizepairs}
C.~Han, J.~Jo, A.~Li, B.~Lee, O.~Deussen, and Y.~Wang.
\newblock Sizepairs: Achieving stable and balanced temporal treemaps using hierarchical size-based pairing.
\newblock {\em IEEE Transactions on Visualization and Computer Graphics}, 29(1):193--202, 2022.

\bibitem{heer2023mosaic}
J.~Heer and D.~Moritz.
\newblock Mosaic: An architecture for scalable \& interoperable data views.
\newblock {\em IEEE Transactions on Visualization and Computer Graphics}, 2023.

\bibitem{holder2022dispersion}
E.~Holder and C.~Xiong.
\newblock Dispersion vs disparity: Hiding variability can encourage stereotyping when visualizing social outcomes.
\newblock {\em IEEE Transactions on Visualization and Computer Graphics}, 29(1):624--634, 2022.

\bibitem{hoque2022visual}
M.~N. Hoque, W.~He, A.~K. Shekar, L.~Gou, and L.~Ren.
\newblock Visual concept programming: A visual analytics approach to injecting human intelligence at scale.
\newblock {\em IEEE Transactions on Visualization and Computer Graphics}, 29(1):74--83, 2022.

\bibitem{isenberg2013systematic}
T.~Isenberg, P.~Isenberg, J.~Chen, M.~Sedlmair, and T.~M{\"o}ller.
\newblock A systematic review on the practice of evaluating visualization.
\newblock {\em IEEE Transactions on Visualization and Computer Graphics}, 19(12):2818--2827, 2013.

\bibitem{komlodi2004information}
A.~Komlodi, A.~Sears, and E.~Stanziola.
\newblock Information visualization evaluation review.
\newblock Technical report, ISRC Tech. Report, Dept. of Information Systems, UMBC, 2004.

\bibitem{lam2011empirical}
H.~Lam, E.~Bertini, P.~Isenberg, C.~Plaisant, and S.~Carpendale.
\newblock Empirical studies in information visualization: Seven scenarios.
\newblock {\em IEEE Transactions on Visualization and Computer Graphics}, 18(9):1520--1536, 2011.

\bibitem{lee2022affective}
E.~Lee-Robbins and E.~Adar.
\newblock Affective learning objectives for communicative visualizations.
\newblock {\em IEEE Transactions on Visualization and Computer Graphics}, 29(1):1--11, 2022.

\bibitem{li2022dual}
Z.~Li, R.~Shi, Y.~Liu, S.~Long, Z.~Guo, S.~Jia, and J.~Zhang.
\newblock Dual space coupling model guided overlap-free scatterplot.
\newblock {\em IEEE Transactions on Visualization and Computer Graphics}, 29(1):657--667, 2022.

\bibitem{linhares2022largenetvis}
C.~D. Linhares, J.~R. Ponciano, D.~S. Pedro, L.~E. Rocha, A.~J. Traina, and J.~Poco.
\newblock Largenetvis: Visual exploration of large temporal networks based on community taxonomies.
\newblock {\em IEEE Transactions on Visualization and Computer Graphics}, 29(1):203--213, 2022.

\bibitem{mcnutt2022no}
A.~M. McNutt.
\newblock No grammar to rule them all: A survey of json-style dsls for visualization.
\newblock {\em IEEE Transactions on Visualization and Computer Graphics}, 29(1):160--170, 2022.

\bibitem{morariu2022predicting}
C.~Morariu, A.~Bibal, R.~Cutura, B.~Fr{\'e}nay, and M.~Sedlmair.
\newblock Predicting user preferences of dimensionality reduction embedding quality.
\newblock {\em IEEE Transactions on Visualization and Computer Graphics}, 29(1):745--755, 2022.

\bibitem{moritz2023average}
D.~Moritz, L.~M. Padilla, F.~Nguyen, and S.~L. Franconeri.
\newblock Average estimates in line graphs are biased toward areas of higher variability.
\newblock {\em IEEE Transactions on Visualization and Computer Graphics}, 2023.

\bibitem{morrical2022quick}
N.~Morrical, A.~Sahistan, U.~G{\"u}d{\"u}kbay, I.~Wald, and V.~Pascucci.
\newblock Quick clusters: A gpu-parallel partitioning for efficient path tracing of unstructured volumetric grids.
\newblock {\em IEEE Transactions on Visualization and Computer Graphics}, 29(1):537--547, 2022.

\bibitem{munzner2008process}
T.~Munzner.
\newblock Process and pitfalls in writing information visualization research papers.
\newblock In {\em Information visualization: human-centered issues and perspectives}, pp. 134--153. Springer, 2008.

\bibitem{munzner2009nested}
T.~Munzner.
\newblock A nested model for visualization design and validation.
\newblock {\em IEEE transactions on visualization and computer graphics}, 15(6):921--928, 2009.

\bibitem{oral2023information}
E.~Oral, R.~Chawla, M.~Wijkstra, N.~Mahyar, and E.~Dimara.
\newblock From information to choice: A critical inquiry into visualization tools for decision making.
\newblock {\em IEEE Transactions on Visualization and Computer Graphics}, 2023.

\bibitem{padilla2022multiple}
L.~Padilla, R.~Fygenson, S.~C. Castro, and E.~Bertini.
\newblock Multiple forecast visualizations (mfvs): Trade-offs in trust and performance in multiple covid-19 forecast visualizations.
\newblock {\em IEEE transactions on visualization and computer graphics}, 29(1):12--22, 2022.

\bibitem{panagiotidou2022communicating}
G.~Panagiotidou, H.~Lamqaddam, J.~Poblome, K.~Brosens, K.~Verbert, and A.~V. Moere.
\newblock Communicating uncertainty in digital humanities visualization research.
\newblock {\em IEEE Transactions on Visualization and Computer Graphics}, 29(1):635--645, 2022.

\bibitem{patil2022studying}
A.~Patil, G.~Richer, C.~Jermaine, D.~Moritz, and J.-D. Fekete.
\newblock Studying early decision making with progressive bar charts.
\newblock {\em IEEE Transactions on Visualization and Computer Graphics}, 29(1):407--417, 2022.

\bibitem{quadri2024do}
G.~J. Quadri, A.~Z. Wang, Z.~Wang, J.~Adorno, P.~Rosen, and D.~A. Szafir.
\newblock Do you see what i see? a qualitative study eliciting high-level visualization comprehension.
\newblock In {\em ACM CHI}, pp. 1--26, 2024.

\bibitem{rodrigues2022relaxed}
N.~Rodrigues, C.~Schulz, S.~D{\"o}ring, D.~Baumgartner, T.~Krake, and D.~Weiskopf.
\newblock Relaxed dot plots: Faithful visualization of samples and their distribution.
\newblock {\em IEEE Transactions on Visualization and Computer Graphics}, 29(1):278--287, 2022.

\bibitem{shen2022idlat}
J.~Shen, H.~Li, J.~Xu, A.~Biswas, and H.-W. Shen.
\newblock Idlat: An importance-driven latent generation method for scientific data.
\newblock {\em IEEE Transactions on Visualization and Computer Graphics}, 29(1):679--689, 2022.

\bibitem{shin2022scanner}
S.~Shin, S.~Chung, S.~Hong, and N.~Elmqvist.
\newblock A scanner deeply: Predicting gaze heatmaps on visualizations using crowdsourced eye movement data.
\newblock {\em IEEE Transactions on Visualization and Computer Graphics}, 29(1):396--406, 2022.

\bibitem{small2011conduct}
M.~L. Small.
\newblock How to conduct a mixed methods study: Recent trends in a rapidly growing literature.
\newblock {\em Annual review of sociology}, 37(1):57--86, 2011.

\bibitem{smart2019color}
S.~Smart, K.~Wu, and D.~A. Szafir.
\newblock Color crafting: Automating the construction of designer quality color ramps.
\newblock {\em IEEE transactions on visualization and computer graphics}, 26(1):1215--1225, 2019.

\bibitem{sperrle2021survey}
F.~Sperrle, M.~El-Assady, G.~Guo, R.~Borgo, D.~H. Chau, A.~Endert, and D.~Keim.
\newblock A survey of human-centered evaluations in human-centered machine learning.
\newblock In {\em Computer Graphics Forum}, vol.~40, pp. 543--568. Wiley Online Library, 2021.

\bibitem{swift2022visualizing}
M.~E. Swift, W.~Ayers, S.~Pallanck, and S.~Wehrwein.
\newblock Visualizing the passage of time with video temporal pyramids.
\newblock {\em IEEE Transactions on Visualization and Computer Graphics}, 29(1):171--181, 2022.

\bibitem{tandon2022measuring}
S.~Tandon, A.~Abdul-Rahman, and R.~Borgo.
\newblock Measuring effects of spatial visualization and domain on visualization task performance: a comparative study.
\newblock {\em IEEE Transactions on Visualization and Computer Graphics}, 29(1):668--678, 2022.

\bibitem{tseng2023evaluating}
C.~Tseng, G.~J. Quadri, Z.~Wang, and D.~A. Szafir.
\newblock Measuring categorical perception in color-coded scatterplots.
\newblock In {\em Proc. ACM Hum. Factors Comput. Syst. (CHI)}, 2023.

\bibitem{vajiac2022trafficvis}
C.~Vajiac, D.~H. Chau, A.~Olligschlaeger, R.~Mackenzie, P.~Nair, M.-C. Lee, Y.~Li, N.~Park, R.~Rabbany, and C.~Faloutsos.
\newblock Trafficvis: Visualizing organized activity and spatio-temporal patterns for detecting and labeling human trafficking.
\newblock {\em IEEE transactions on visualization and computer graphics}, 29(1):53--62, 2022.

\bibitem{wall2022vishikers}
E.~Wall, C.~Xiong, and Y.-S. Kim.
\newblock Vishikers’ guide to evaluation: Competing considerations in study design.
\newblock {\em IEEE Computer Graphics and Applications}, 42(3):29--38, 2022.

\bibitem{warchol2022visinity}
S.~Warchol, R.~Krueger, A.~J. Nirmal, G.~Gaglia, J.~Jessup, C.~C. Ritch, J.~Hoffer, J.~Muhlich, M.~L. Burger, T.~Jacks, et~al.
\newblock Visinity: Visual spatial neighborhood analysis for multiplexed tissue imaging data.
\newblock {\em IEEE transactions on visualization and computer graphics}, 29(1):106--116, 2022.

\bibitem{yu2022pseudo}
Y.~Yu, D.~Kruyff, J.~Jiao, T.~Becker, and M.~Behrisch.
\newblock Pseudo: Interactive pattern search in multivariate time series with locality-sensitive hashing and relevance feedback.
\newblock {\em IEEE Transactions on Visualization and Computer Graphics}, 29(1):33--42, 2022.

\bibitem{yuan2022visual}
J.~Yuan, M.~Liu, F.~Tian, and S.~Liu.
\newblock Visual analysis of neural architecture spaces for summarizing design principles.
\newblock {\em IEEE Transactions on Visualization and Computer Graphics}, 29(1):288--298, 2022.

\bibitem{zong2022animated}
J.~Zong, J.~Pollock, D.~Wootton, and A.~Satyanarayan.
\newblock Animated vega-lite: Unifying animation with a grammar of interactive graphics.
\newblock {\em IEEE Transactions on Visualization and Computer Graphics}, 29(1):149--159, 2022.

\end{thebibliography}

\end{document}